\documentclass[aps,prl,showpacs,amsmath,amssymb,amsfonts,superscriptaddress,twocolumn,lengthcheck]{revtex4-1}

\usepackage{graphicx}
\usepackage{subfigure}
\usepackage{verbatim}
\usepackage{dcolumn}
\usepackage{bm}
\usepackage{epsf}
\usepackage{xcolor}
\usepackage{hyperref}
\usepackage{hhline}
\usepackage{float}
\usepackage{enumerate}
\usepackage{bbm}
\usepackage{lipsum}
\definecolor{mygreen}{rgb}{0.01, 0.31, 0.59}
\definecolor{myblue}{rgb}{0.01, 0.31, 0.59}
\hypersetup{
	colorlinks=true,   
	linkcolor=blue,  
	citecolor=blue, 
	urlcolor =blue    
}

\begin{document}

\title{Entanglement Emerges from Dissipation-Structured Quantum Self-Organization}
\author{Zhi-Bo Yang}
\affiliation{Interdisciplinary Center of Quantum Information, State Key Laboratory of Modern Optical Instrumentation, and Zhejiang Province Key Laboratory of Quantum Technology and Device, Department of Physics, Zhejiang University, Hangzhou 310027, China}

\author{Yi-Pu Wang}
\email{yipuwang@zju.edu.cn}
\affiliation{Interdisciplinary Center of Quantum Information, State Key Laboratory of Modern Optical Instrumentation, and Zhejiang Province Key Laboratory of Quantum Technology and Device, Department of Physics, Zhejiang University, Hangzhou 310027, China}

\author{Jie Li}
\email{jieli007@zju.edu.cn}
\affiliation{Interdisciplinary Center of Quantum Information, State Key Laboratory of Modern Optical Instrumentation, and Zhejiang Province Key Laboratory of Quantum Technology and Device, Department of Physics, Zhejiang University, Hangzhou 310027, China}

\author{C.-M. Hu}
\affiliation{Department of Physics and Astronomy, University of Manitoba, Winnipeg, Canada R3T 2N2}

\author{J. Q. You}
\email{jqyou@zju.edu.cn}
\affiliation{Interdisciplinary Center of Quantum Information, State Key Laboratory of Modern Optical Instrumentation, and Zhejiang Province Key Laboratory of Quantum Technology and Device, Department of Physics, Zhejiang University, Hangzhou 310027, China}

\date{\today}
\begin{abstract}
Entanglement is a holistic property of multipartite quantum systems, which is accompanied by the establishment of nonclassical correlations between subsystems. Most entanglement mechanisms can be described by a coherent interaction Hamiltonian, and entanglement develops over time. In other words, the generation of entanglement has a time arrow. Dissipative structure theory directs the evolving time arrow of a non-equilibrium system. By dissipating energy to the environment, the system establishes order out of randomness. This is also referred to as self-organization. Here, we explore a new mechanism to create entanglement, utilizing the wisdom of dissipative structure theory in quantum systems. The entanglement between subsystems can emerge via the dissipation-structured correlation. This method requires a non-equilibrium initial state and cooperative dissipation, which can be implemented in a variety of waveguide-coupled quantum systems.
\end{abstract}
\maketitle


\textit{Introduction.---}The time arrow and irreversible processes in physics are closely related to the formulation of the second law of thermodynamics, which asserts that the entropy of a closed system does not decrease. The system's entropy will reach its maximum at the stable state of thermodynamic equilibrium, which is delimited by a loose boundary. The dissipative structure theory~\cite{Prigogine-67,Prigogine-78} extends the second law of thermodynamics to non-equilibrium systems, settling the boundary to the state of ``least dissipation"~\cite{Kondepudi-15}. The steady states correspond to a minimum of entropy production in linear non-equilibrium thermodynamics. The dissipative structures describe a novel type of dynamic states of matter, where \textit{dissipative} processes create and maintain the self-organized non-equilibrium structures. 

Under the \textit{constructive} role of irreversible processes, non-equilibrium dynamics become a source of order. However, non-equilibrium is required but not sufficient for the formation of dissipative structure and, ultimately, self-organization~\cite{Prigogine-78}. The critical condition is that the dissipations of the subsystems are synergic rather than disordered. This mechanism can be found in the classical world, such as Huygens pendulum synchronization, wherein two pendulums dissipate energy to the same house beam~\cite{Huygens,Huygens-2,Hu-21}. The fireflies synchronize by copying other fireflies around them, as the result of all the fireflies glowing to the same environment. Synchronizations in this context are unambiguous self-organization behaviors. Such mechanisms can also be found in more complex systems, e.g., organism dynamics~\cite{Levine-10} and social economic activities~\cite{Foster-97}.

Time arrow also exists in the quantum world. A key question is how dissipative structure theory's wisdom will behave in a quantum system. It will be revealed that non-equilibrium dynamics and dissipation can also create a new order in the quantum system. \textit{Quantum} self-organization leads to \textit{entanglement}. Two factors are required: the system should be perturbed to become non-equilibrium and the dissipations of the subsystems should be synergic. Cooperative dissipation has been studied in a variety of systems and in many aspects, including Friedrich-Wintgen bound state in the continuum~\cite{FW}, subradiance~\cite{Sub-1,Sub-2}, decoherence-free subspaces~\cite{Duan,Lidar}, vacuum-induced coherence~\cite{VIC-1,VIC-2}, and cancellation of spontaneous emission~\cite{ZhuScully-96}. They are all related to the destructive interference in radiative decay channels and a ``least dissipation" state can be observed. This cooperative dissipation effect has recently been termed as dissipative couplings~\cite{Clerk-15,Xiao-16,Harder-18,Wang-20,Agarwal-21}. Previous studies have primarily been focused on equilibrium issues and steady states of the dissipatively coupled systems. The dynamic property of quantum state evolution and self-organization behavior remain underexplored.


In this Letter, we investigate the non-equilibrium state (NES) evolution of qubits dissipating cooperatively to a common reservoir. In the \textit{ideal} case of two resonant qubits, we first perturb the system to a NES, where the total decoherence processes of the two individual qubits are synergic, and the system finally evolves to an entangled mixed state. The two-qubit system can no longer radiate, and cooperative dissipation drives the system to enter the ``least dissipation" state. The entanglement is the result of dissipation-structured self-organization. Then we consider the situation where the dissipations are non-synergic and the two qubits have some detuning. We see that the system tends to establish entanglement, but the damping factors eventually destroy the quantum correlation. Correspondingly, we solve this problem by introducing a {\it quantum clockwork} (uninterrupted pump) into the system. The entanglement can be sustained. Further, we extend the mechanism to a three-qubit system. As expected, self-organized entanglement still emerges, because the systems described by dissipative structure theory are generally large ensembles. The quantum dissipative structure presents a novel method to entangle a large number of qubits. Extending dissipation-based self-organization to the quantum region opens up new research directions.

\begin{figure}[t]
	\hskip-0.182cm\includegraphics[width=\linewidth]{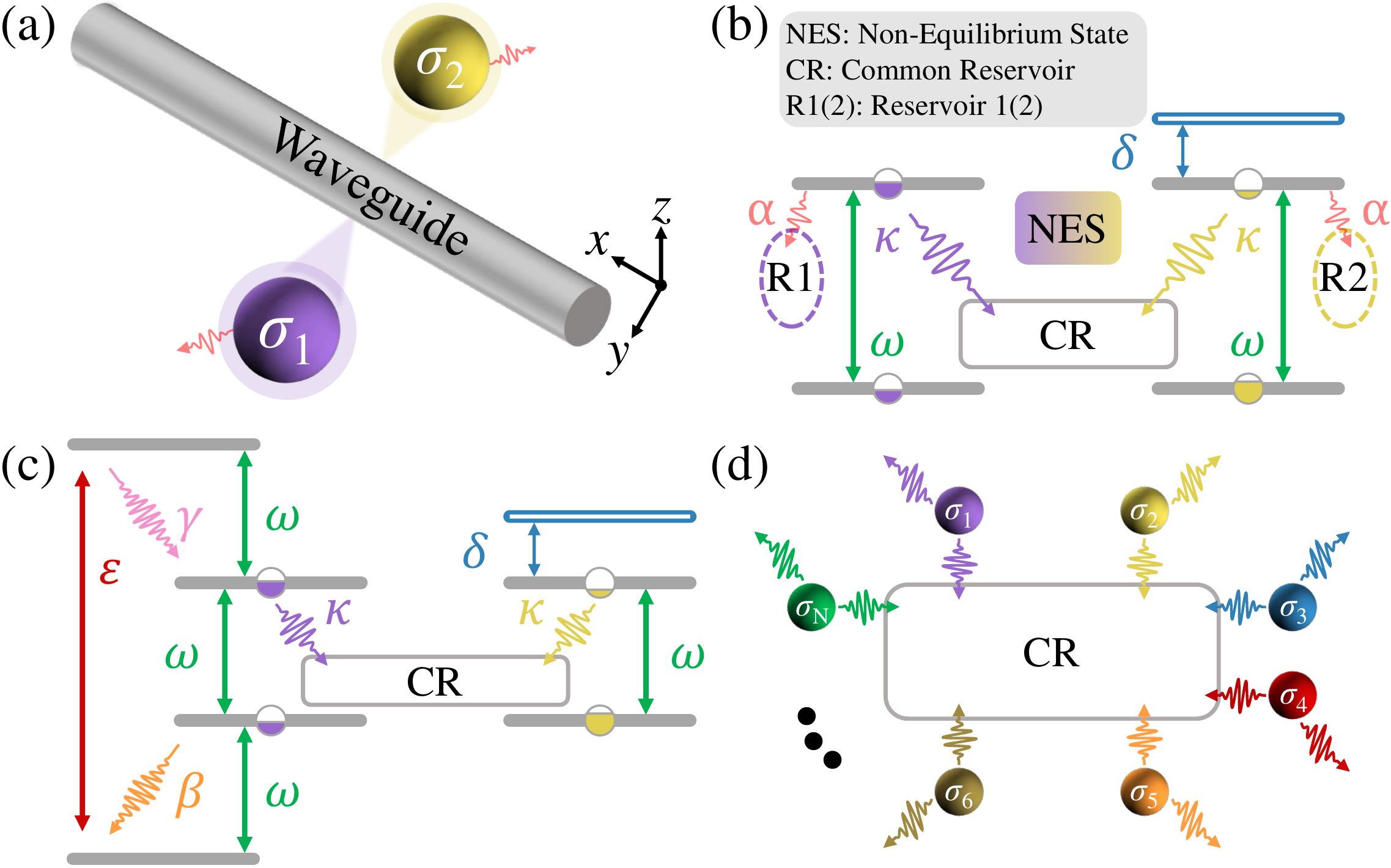}
	\caption{(a) Sketch of the cooperative dissipation system. Two qubits are placed near a common reservoir (waveguide), which supports qubits to dissipate energy cooperatively. (b) Schematic diagram of the entangled state generation mechanism. A non-equilibrium initial state is needed. The ideal case corresponds to two identical qubits that cooperatively decay to the common reservoir (CR) only. In the more general case, two detuned qubits decay to the common reservoir and also their respective reservoirs 1 and 2 (R1 and R2). (c) Schematic diagram of {\it quantum} {\it clockwork} configuration,  which provides the system with an uninterrupted pump. (d) An illustration of the {\it N}-qubit cooperative dissipation system.}
	\label{fig1}
\end{figure}

\textit{Model.---}The essential system is represented schematically in Fig.~\ref{fig1}(a) as waveguide-coupled quantum emitters, where two qubits dissipate to a zero-temperature common reservoir (e.g., a waveguide). Simultaneously, the two qubits can also dissipate to their own reservoirs, which is attributed to the non-cooperative dissipations. When only considering the cooperative dissipation, the dissipative system can be described by the following Lindblad master equation 
\begin{eqnarray}\label{e001}
	\frac{d}{dt}\rho&=&-\frac{i}{\hbar}[H,\rho]+\kappa\mathcal{L}[o^{-}]\rho,
\end{eqnarray}
where $H/\hbar=\omega\sigma_1^+\sigma_1^-+(\omega+\delta)\sigma_2^+\sigma_2^-$ denotes the free Hamiltonian of the two qubits with excitation frequency $\omega$ and detuned frequency $\delta$. The two qubits have \textit{no} direct interaction, and they have the same decay rate $\kappa$. Here $\sigma_{1,2}^+$ ($\sigma_{1,2}^-$) is the Pauli raising (lowering) operator. $\mathcal{L}[o^{-}]$ is the standard dissipative superoperator, and it can be expanded as follows:
\begin{eqnarray}\label{e002}
	\mathcal{L}[o^{-}]\rho=2o^{-}\rho o^{+}-o^{+}o^{-}\rho-\rho o^{+}o^{-},
\end{eqnarray}
where the jump operator needs to take the form $o^{-}=\sigma_1^-+\sigma_2^-$ for depicting the cooperative dissipation~\cite{Carmichael-93}.

\textit{Entanglement emerges from cooperative dissipations.---}Our approach starts with the \textit{ideal} case of two resonant qubits, we first explore the eigenstates and eigenvalues, and then analyse the state and energy evolution from a non-equilibrium initial state. In the four-dimensional Hilbert space, and under the resonance condition ($\delta=0$), the effective Hamiltonian is written as 
\begin{eqnarray}\label{e003}
H/\hbar&=&(\omega-2i\kappa)\vert\Psi_{+}\rangle\langle\Psi_{+}\vert+\omega\vert\Psi_{-}\rangle\langle\Psi_{-}\vert\nonumber\\
&&+2(\omega-i\kappa)\vert11\rangle\langle11\vert
\end{eqnarray}
with $\vert\Psi_{\pm}\rangle=(\vert10 \rangle\pm\vert01 \rangle)/\sqrt{2}$~\cite{supp}. We can see that: (i) $\vert\Psi_{+}\rangle$, the symmetric Bell state, is a bright state that will decay at twice the decay rate due to the superradiance effect; and (ii) $\vert\Psi_{-}\rangle$, the anti-symmetric Bell state, is a dark state, which will not dissipate due to the subradiance effect. There are entangled states in the eigenstates of the dissipatively coupled system and one of them is the ``least dissipation" state, which essentially lays out the main point of this Letter. Next, we need to investigate whether the system can evolve towards the corresponding ``least dissipation" state from different initial states, and then finally establish entanglement.

\begin{figure}[t]
	\hskip-0.19cm\includegraphics[width=\linewidth]{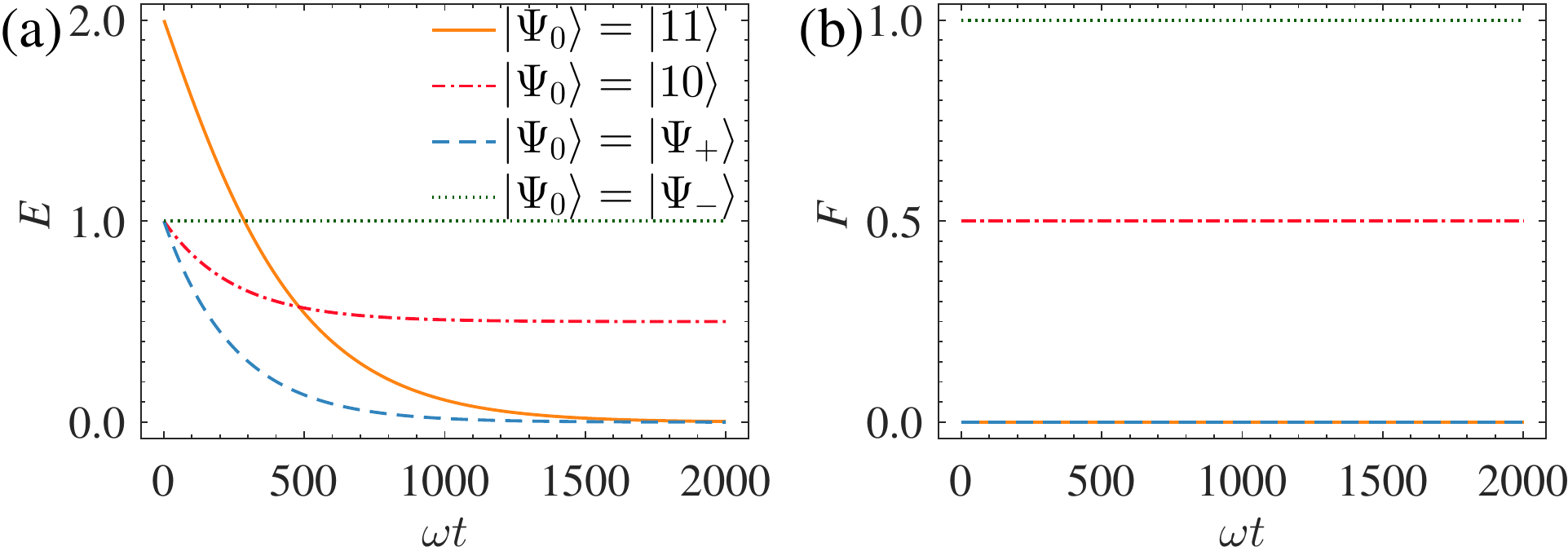}
	\caption{(a) Energy {\it E} and (b) Fidelity {\it F} versus evolution time $\omega t$ for four different initial states, where we set $\omega=1$ and $\kappa/\omega=0.001$.}
	\label{fig2}
\end{figure}

The evolution of energy (in units of $\hbar\omega$) from four different initial states ($\vert11 \rangle$, $\vert10 \rangle$, $\vert\Psi_{-}\rangle$, $\vert\Psi_{+}\rangle$) are calculated by using Eq.~(\ref{e001})~\cite{supp} and plotted in Fig.~\ref{fig2}(a). Looking at the eigenvalues, it is clear that the $\vert11 \rangle$ and $\vert\Psi_{+}\rangle$ will eventually dissipate all of the energy, whereas the dark state $\vert\Psi_{-}\rangle$ will never dissipate. The most fascinating thing happens when the initial state is a NES, which is $\vert10\rangle$. After dissipating half of the energy, the two-qubit system will become stable. This indicates that the system finally evolves to the ``least dissipation" state. To see the detailed state evolution, we monitor the state fidelity with respect to the dark state. The fidelity is defined as {\it F} $=(\vert\Psi_{-}\rangle$, $\rho)=tr\sqrt{\langle\Psi_{-}\vert\rho\vert\Psi_{-}\rangle}$. {\it F} as a function of evolution time $\omega t$ is shown in Fig.~\ref{fig2}(b). We can find that the fidelity of evolution states from initial states $\vert11 \rangle$ and $\vert\Psi_{+}\rangle$ relative to state $\vert\Psi_{-}\rangle$ is always zero. This implies that states $\vert11 \rangle$ and $\vert\Psi_{+}\rangle$ will never evolve to the dark state. While the states revolting from $\vert10 \rangle$ and $\vert\Psi_{-}\rangle$, the fidelities relative to state $\vert\Psi_{-}\rangle$ are always 0.5 and 1, respectively. It is worth noting that state $\vert10 \rangle$ can be written as a superposition state $(\vert\Psi_{+}\rangle+\vert\Psi_{-}\rangle)/\sqrt{2}$. During the evolution process, $\vert\Psi_{+}\rangle/\sqrt{2}$ will evolve to the ground state, so the fidelity of state $\vert10 \rangle$ is always $0.5$. That is, the final state is $\rho_{2}(\infty)=(\vert\Psi_{-}\rangle\langle\Psi_{-}\vert+\vert00\rangle\langle00\vert)/2$, which is an entangled mixed state~\cite{supp,Peters-PRL,Wei-PRA}. 

The main distinction between the $\vert01\rangle$ state and the others is that it is a NES with unequal excitation probabilities for the two qubits. The physics behind this is that any NES contains a dark state component. For two qubits $A$ and $B$, we can have $\rho_{A}=a\vert0\rangle\langle0\vert+b\vert1\rangle\langle1\vert$, $\rho_{B}=c\vert0\rangle\langle0\vert+d\vert1\rangle\langle1\vert$, $a+b=c+d=1$, and $\rho_{AB}=ac\vert00\rangle\langle00\vert+bd\vert11\rangle\langle11\vert+2(bc-ad)\vert\Psi_{-}\rangle\langle\Psi_{-}\vert+2(ad+bc)\vert\Psi_{+}\rangle\langle\Psi_{+}\vert$. If $b\neq d$, then $2(bc-ad)\neq 0$. As a result, we have a two-qubit NES, and it clearly indicates the presence of the dark-state component. All non-dark states will be dissipated as the system evolves, resulting in the final entangled mixed state. This finding enriches the classical dissipative structure theory. The presence of non-equilibrium indicates the existence of the dark-state component, which is why the non-equilibrium is the source of order. We can now say that non-equilibrium \textit{contains} order. The most significant advantage of the cooperative dissipation system is that once the entangled state is established, it will not be decohered, which will greatly increase the coherence time, especially when continuous pumping is lacking.

{\it Non-ideal cases and quantum clockwork}.---This section focuses on the entanglement decay caused by non-cooperative dissipations and detuning between qubits. As depicted in Fig.~\ref{fig1}(b), the individual dissipations of the qubits to their respective reservoirs 1 and 2 are attributed to the non-cooperative dissipations $\alpha$, which can be introduced into the system by inserting the standard Lindblad operator $\sum_{j=1}^{2}\alpha\mathcal{L}[\sigma^{-}_j]\rho$ into Eq.~(\ref{e001}). Then we introduce a four-level qudit to provide auxiliary energy levels for loading the pump, which does not disturb the cooperative dissipation evolved in the entanglement generation. For the cooperative dissipations $\kappa$, they only play a role in the decay of non-dark states as shown in Fig.~\ref{fig2}(a). However, non-cooperative dissipations $\alpha$ have an effect not only on the decay of the non-dark state, but also on the decay of the dark state, shifting the ``least dissipation'' state from zero dissipation one to finite dissipation one. Furthermore, the frequency detuning $\delta$ between the qubits will break the dark state, resulting in persistent state leaking.

\begin{figure}[t]
	\hskip-0.18cm\includegraphics[width=\linewidth]{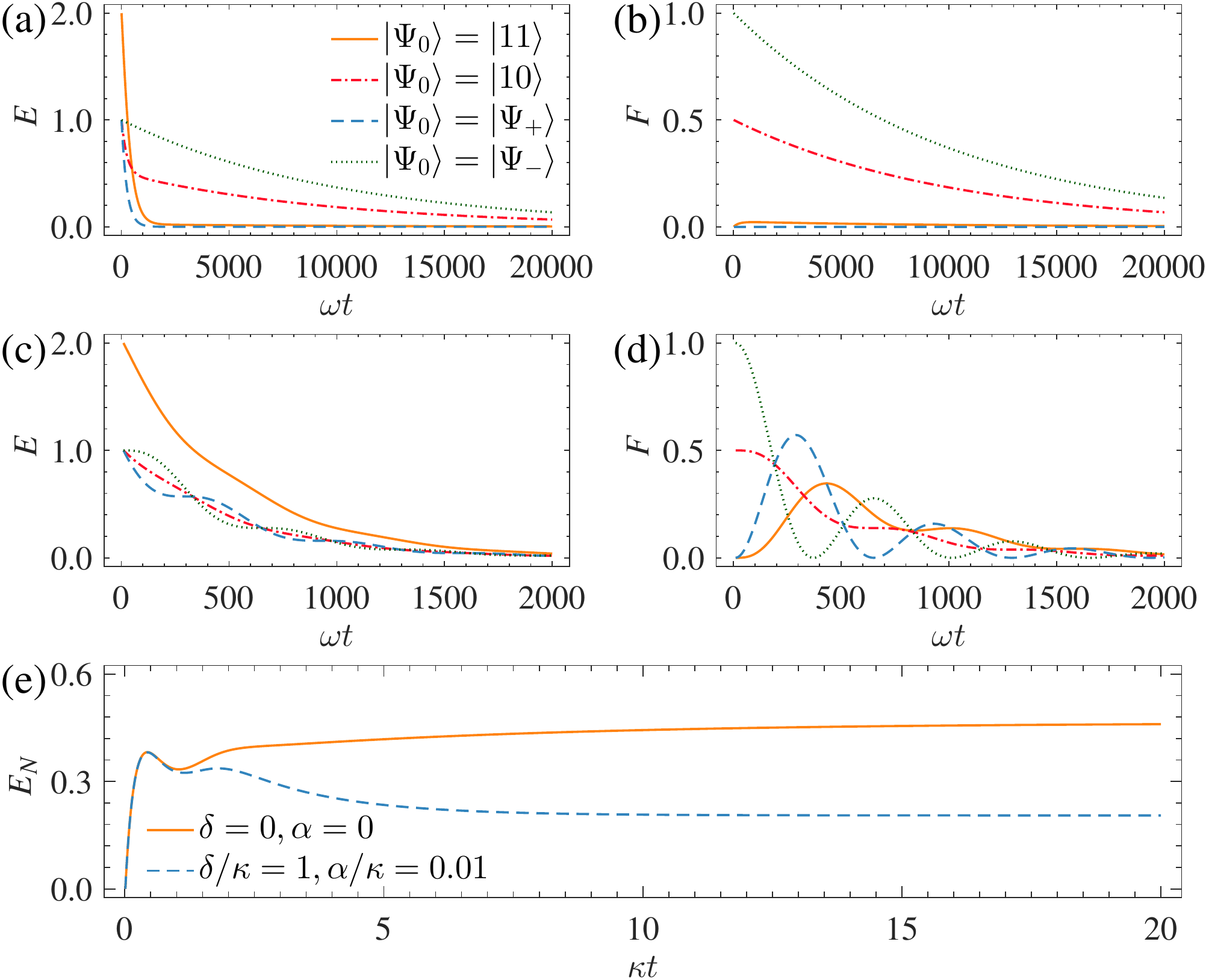}
	\caption{(a) (c) Energy {\it E} and (b) (d) Fidelity {\it F} versus evolution time $\omega t$ for four initial states, with $\alpha/\kappa=0.05$, $\delta=0$ in (a) (b) and $\alpha=0$, $\delta/\omega=0.1$ in (c) (d) and the other parameters are $\omega=1$, $\kappa/\omega=0.001$. (e) Logarithmic negativity $E_N$ versus evolution time $\kappa t$ with $\kappa=1$, $\gamma/\kappa=3$, $\beta/\kappa=0.1$, and $\varepsilon/\kappa=2$. The corresponding configuration is depicted in Fig.~\ref{fig1}(c).}
	\label{fig3}
\end{figure}

In order to show the effects of non-cooperative dissipations and detuning on the entangled state, we solve the dynamic equation and track the time evolution of both energy and fidelity. The effects of non-cooperative dissipation $\alpha$ on the energy and state evolutions are depicted in Figs.~\ref{fig3}(a) and \ref{fig3}(b), respectively. We can see that if the initial state is a dark state, the energy and fidelity will only be affected by non-cooperative dissipation, leading to a smooth exponential decay (green dotted curve). When the initial state is a NES $\vert10\rangle$, the reduction of fidelity is only attributed to the dark state decay. However, due to the synthetic cooperative and non-cooperative dissipations, the energy will quickly decay to $0.5~\hbar\omega$  (red dot-dashed curve). At this point, most of the non-dark state component is exhausted, and the residual part is mainly the dark state. An entangled mixed state is formed. Then the non-cooperative dissipation further decays the state to the ground state slowly. In this case, the system tends to reach a metastable state (i.e., an entangled mixed state) before it reaches the ground state. If the initial state is an equilibrium excited state $\vert11\rangle$ or bright state, the energy will rapidly decay to zero, as shown by the orange solid and blue dashed curves. Since NES is the necessity of a dissipative structure, we can conclude that the energy of a quantum system decays at a faster rate than would occur if the dissipative structure did not exist.

Setting $\alpha=0$, we investigate the entanglement and energy decay caused by frequency detuning $\delta$, as depicted in Figs.~\ref{fig3}(c) and \ref{fig3}(d). We find that even without non-cooperative dissipation, the energy will eventually be exhausted regardless of the system's initial state. Indeed, the presence of detuning breaks the dark state, resulting in both a decay factor and a phase factor of the ``least dissipation" eigenstate~\cite{supp}. This causes the fidelity and energy to rapidly decay to zero with periodic oscillations.

In the pendulum synchronization experiment, due to the intrinsic friction loss, the synchronized pendulums will eventually stop moving, corresponding to the ground state of the quantum system. The similarity between the classical and quantum systems is that before the energy is completely dissipated, self-organization driven by non-equilibrium will cause the classical and quantum systems to enter the ``least dissipation" state. It is the synchronized state in the classical case, and the entangled state in the quantum case. Interestingly, a clockwork can be added to provide continuous power to the classical pendulum without affecting its normal swing period. Drawing on this wisdom, we replace one qubit with a four-level qudit and pump the corresponding transition, so as to provide a continuous working energy source for the evolution of quantum self-organization, i.e., introducing the idea of {\it quantum clockwork}, as schematically shown in Fig.~\ref{fig1}(c). See supplementary material for more details~\cite{supp}. While pumping the system, we can monitor the entanglement evolution of the qubit-qudit system. The entanglement can be characterized by the logarithmic negativity $E_N=\log_{2}\Vert\rho^{T}\Vert_1$, where the symbol $\Vert$ $\Vert_1$ denotes the trace norm, and $\rho^{T}$ is the partial transpose of the reduced density matrix $\rho$ of the two sub-systems~\cite{EN}. The logarithmic negativity $E_N$ is plotted as a function of time $\kappa t$ in Fig.~\ref{fig3}(e), demonstrating that the steady-state entanglement can be generated with the help of quantum clockwork, even when the non-cooperative dissipation and detuning exist.


\textit{Three-qubit W-like state generation.---}Dissipative structures in nature typically contain many subsystems. After a long evolution time, self-organization is finally established. Quantum system has no exception. Starting from NES, we can establish multi-qubit quantum entanglement in principle through cooperative dissipation. Next, we will take a three-qubit system as an example to discuss the process of building the \textit{W}-like entangled state in detail, and discuss the possibility of extending it to \textit{N}-qubit systems. In this section, we consider the ideal case where the properties of the three qubits are exactly the same and the non-cooperative dissipations are zero. 

We can find that any single-excitation NES can be represented by basis vectors $\vert\Psi_{1}\rangle=(\vert 100\rangle+\vert 010\rangle+\vert 001\rangle)/\sqrt{3}$, $\vert\Psi_{2}\rangle=(2\vert100\rangle-\vert010\rangle-\vert001\rangle)/\sqrt{6}$, and $\vert\Psi_{3}\rangle=(\vert010\rangle-\vert 001\rangle)/\sqrt{2}$, where $\vert\Psi_{2}\rangle$ and $\vert\Psi_{3}\rangle$ are dark states of the system. For example, the non-equilibrium initial state $\vert100\rangle$ can be written as $\vert100\rangle=\vert\Psi_{1}\rangle/\sqrt{3}+2\vert\Psi_{2}\rangle/\sqrt{6}$. However, self-organizing behavior will dissipate all non-dark states. Therefore, the final state of the system is $\rho_3(\infty)=(2\vert\Psi_{2}\rangle\langle\Psi_{2}\vert+\vert000\rangle\langle000\vert)/3$, which is a {\it W}-like entangled mixed state~\cite{supp}. In the dynamical illustration, the photon dissipated by qubit 1 enters the common reservoir, and this photon may be absorbed by either qubit 2 or qubit 3. This process realizes the entanglement among the three qubits. It should be noted that when the non-equilibrium initial state is $\vert 011 \rangle$, since all the double-excitation basis vectors are non-dark states~\cite{supp}, the system will further decay and keep evolving to the ``least dissipation'' state $\rho_3(\infty)$~\cite{supp}. 

\begin{figure}[t]
	\hskip-0.18cm\includegraphics[width=\linewidth]{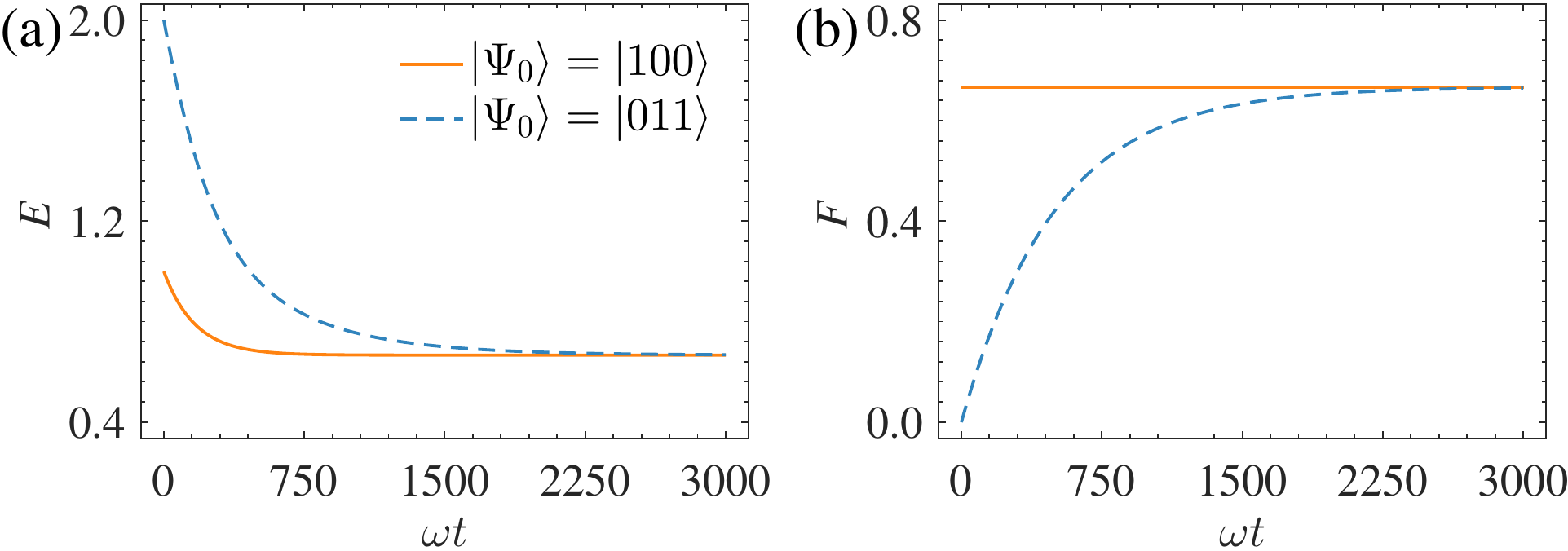}
	\caption{(a) Energy {\it E} and (b) Fidelity {\it F} versus evolution time $\omega t$ for two initial states, where we take $\omega=1$ and $\kappa/\omega=0.001$.}
	\label{fig4}
\end{figure}

From different initial states ($\vert 100 \rangle$ and $\vert 011 \rangle$), the energies {\it E} as a function of time are plotted in Fig.~\ref{fig4}(a). When the system establishes three-qubit entanglement through cooperative dissipation, the system will definitely evolve to the state $\rho_3(\infty)$. In this process, all non-dark states will decay to the ground state $\vert 000 \rangle$, and approximately $0.67~\hbar\omega$ energy will remain, which is the intrinsic energy of the dark state part $2\vert\Psi_{2}\rangle/\sqrt{6}$ of the initial state. The fidelities versus the dark state $\vert \Psi_2 \rangle$ as a function of time are plotted in Fig.~\ref{fig4}(b). The $\vert 100 \rangle$ state contains a certainly proportion of the dark state $\vert \Psi_2 \rangle$, and $\vert \Psi_2 \rangle$ will not decay. Therefore, the fidelity is invariable. For the initial state $\vert 011 \rangle$, the excitation number is 2. At $\omega t=0$, the fidelity versus the single-excitation state $\vert \Psi_2 \rangle$ is zero. As the system dissipates energy, it finally evolves to $\rho_3(\infty)$, and the fidelity reaches the same value as that in the $\vert 100 \rangle$ case.


Finally, the non-equilibrium {\it N}-qubit cooperative dissipation system can also generate the {\it W}-like state through the quantum self-organization, which is schematically shown in Fig.~\ref{fig1}(d). The cooperative dissipation system composed of {\it N} identical qubits with non-equilibrium initial state can evolve to the {\it W}-like entangled mixed state $\rho_{N}(\infty)$ (least dissipation state), which only contains the dark-state components and the {\it N}-qubit ground state. Also, the relative phases between the dissipations of the qubits should be considered, which is closely related to the relative position between the qubits in the waveguide setup.

\textit{Conclusion and perspectives.---}Dissipation is not necessarily a detrimental occurrence. In this work, we study the framework of dissipative structures in the quantum world. The key finding is that non-equilibrium and cooperative dissipation can drive the quantum self-organization. The second law of thermodynamics specifies which processes are allowed to occur in nature, such as the requirement of nondecreasing entropy in an isolated system. It predicts that everything will eventually approach the maximum entropy state. The dissipative structure theory, on the other hand, tells which processes tend to occur. An example is that if the system is perturbed, the entropy production will increase, but the system reacts by returning to the minimum value of the entropy production state. This state is also the ``least dissipation'' state. By introducing this wisdom to the quantum world, we find that a quantum system also tends to ``drop'' into the quantum ``least dissipation'' state. Profoundly, this state could be an entangled mixed state, which is unique to quantum systems when compared with classical systems. Our work reveals the tip of the iceberg of quantum self-organization behavior, and will open up a new direction for studying dissipative structure theory in quantum systems.

We also discuss the new entanglement generation mechanism in detail, and study the \textit{ideal} and \textit{non-ideal} cases, two-qubit and three-qubit models, and the possibility of extending it to \textit{N}-qubit systems. The scheme could be implemented in waveguide-coupled ultracold atoms~\cite{Vetsch-PRL,Corzo-19}, superconducting qubits~\cite{Gu-17,Painter-19}, or hybrid quantum emitters and circuits~\cite{Elshaari-20,Kim-17,Nori-13}.

\begin{acknowledgments}
This work is supported by the National Natural Science Foundation of China (Nos.~11934010, U1801661 and 11774022), the National Key Research and Development Program of China (No.~2016YFA0301200), Zhejiang Province Program for Science and Technology (No.~2020C01019), and the Fundamental Research Funds for the Central Universities (No.~2021FZZX001-02). C.-M. H. is funded by the NSERC Discovery Grants and the NSERC Discovery Accelerator Supplements.
\end{acknowledgments}

\appendix


\begin{thebibliography}{99}
\bibitem{Prigogine-67}	
I. Prigogine, \emph{Introduction to Thermodynamics of Irreversible Processes} (John Wiley \& Sons, New York, 1967).

\bibitem{Prigogine-78}
I. Prigogine, Time, Structure, and Fluctuations, Science {\bf 201}, 777 (1978).

\bibitem{Kondepudi-15}	
D. Kondepudi and I. Prigogine, \emph{Modern thermodynamics: from heat engines to dissipative structures} (John Wiley \& Sons, 2015).

\bibitem{Huygens}
C. Huygens, \emph{Horologium Oscillatorium}, translated by R. J. Blackwell (Iowa
State University Press, 1986). 

\bibitem{Huygens-2}
H. M. Oliveira and L. V. Melo, Huygens synchronization of two clocks, Scientific Reports {\bf 5}, 11548 (2015).

\bibitem{Hu-21}
M. Harder, B. M. Yao, Y. S. Gui, and C.-M. Hu, Coherent and dissipative cavity magnonics, Journal of Applied Physics {\bf 129}, 201101 (2021).


\bibitem{Levine-10}
E. Ben-Jacob, I. Cohen, H. Levine, Cooperative self-organization of microorganisms, Advances in Physics {\bf 49}, 395 (2000).

\bibitem{Foster-97}
J. Foster, The analytical foundations of evolutionary economics: From biological analogy to economic self-organization, Structural change and economic dynamics {\bf 8}, 427 (1997).
	
\bibitem{FW}
H. Friedrich and D. Wintgen, Interfering resonances and bound states in the continuum, Phys. Rev. A {\bf 32}, 3231 (1985).

\bibitem{Sub-1}
H. Freedhoff and J. van Kranendonk, Theory of coherent resonant absorption and emission at infrared and optical frequencies, Can. J. Phys. {\bf 45}, 1833 (1967).

\bibitem{Sub-2}
W. Guerin, M. O. Araújo, and R. Kaiser, Subradiance in a Large Cloud of Cold Atoms, Phys. Rev. Lett. {\bf 116}, 083601 (2016).

\bibitem{Duan}
L.-M. Duan and G.-C. Guo, Preserving Coherence in Quantum Computation by Pairing Quantum Bits, Phys. Rev. Lett. {\bf 79}, 1953 (1997).

\bibitem{Lidar}
D. A. Lidar, I. L. Chuang, and K. B. Whaley, Decoherence-Free Subspaces for Quantum Computation, Phys. Rev. Lett. {\bf 81}, 2594 (1998).

\bibitem{VIC-1}
G. S. Agarwal, Anisotropic Vacuum-Induced Interference in Decay Channels, Phys. Rev. Lett. {\bf 84}, 5500 (2000).

\bibitem{VIC-2}
P. K. Jha, X. Ni, C. Wu, Y. Wang, and X. Zhang, Metasurface-Enabled Remote Quantum Interference, Phys. Rev. Lett. {\bf 115}, 025501 (2015).

\bibitem{ZhuScully-96}
S.-Y. Zhu, and M. O. Scully, Spectral Line Elimination and Spontaneous Emission Cancellation via Quantum Interference, Phys. Rev. Lett. {\bf 76}, 388 (1996).

\bibitem{Clerk-15}
A. Metelmann and A. A. Clerk, Nonreciprocal Photon Transmission and Amplification via Reservoir Engineering, Phys. Rev. X {\bf 5}, 021025 (2015).

\bibitem{Xiao-16}
P. Peng, W. Cao, C. Shen, W. Qu, J. Wen, L. Jiang, and Y. Xiao, Anti-parity-time symmetry with flying atoms, Nature Physics {\bf 12}, 1139 (2016).

\bibitem{Harder-18}
M. Harder, Y. Yang, B. M. Yao, C. H. Yu, J. W. Rao, Y. S. Gui, R. L. Stamps, and C.-M. Hu, Level Attraction Due to Dissipative Magnon-Photon Coupling, Phys. Rev. Lett. {\bf 121}, 137203 (2018).

\bibitem{Wang-20}
Y.-P. Wang and C.-M. Hu, Dissipative couplings in cavity magnonics, Journal of Applied Physics {\bf 127}, 130901 (2020).

\bibitem{Agarwal-21}
J. M. P. Nair, D. Mukhopadhyay, and G. S. Agarwal, Enhanced Sensing of Weak Anharmonicities through Coherences in Dissipatively Coupled Anti-PT Symmetric Systems, Phys. Rev. Lett. {\bf 126}, 180401 (2021).

\bibitem{Carmichael-93}
H. J. Carmichael, Quantum Trajectory Theory for Cascaded Open Systems, Phys. Rev. Lett. {\bf 70}, 2273 (1993).

\bibitem{supp}
See Supplementary Material at ... for additional details, which includes Ref.~\cite{Carmichael-93}.

\bibitem{Wei-PRA}
T.-C. Wei, K. Nemoto, P. M. Goldbart, P. G. Kwiat, W. J. Munro, and F. Verstraete, Maximal entanglement versus entropy for mixed quantum states, Phys. Rev. A {\bf 67}, 022110 (2003).

\bibitem{Peters-PRL}
N. A. Peters, J. B. Altepeter, D. Branning, E. R. Jeffrey, T.-C. Wei, and P. G. Kwiat, Maximally Entangled Mixed States: Creation and Concentration, Phys. Rev. Lett. {\bf 92}, 133601 (2004).

\bibitem{EN}
G. Vidal and R. F. Werner, Computable measure of entanglement, Phys. Rev. A \textbf{65}, 032314 (2002).

\bibitem{Vetsch-PRL}
E. Vetsch, D. Reitz, G. Sagué, R. Schmidt, S. T. Dawkins, A. Rauschenbeutel, Optical interface created by laser-cooled atoms trapped in the evanescent field surrounding an optical nanofiber, Phys. Rev. Lett. {\bf 104}, 203603 (2010).

\bibitem{Corzo-19}
N. V. Corzo, J. Raskop, A. Chandra, A. S. Sheremet, B. Gouraud, and J. Laurat, Waveguide-coupled single collective excitation of atomic arrays, Nature Physics {\bf 566}, 359 (2019).

\bibitem{Gu-17}
X. Gu, A. F. Kockum, A. Miranowicz, Y.-x. Liu, and F. Nori, Microwave photonics with superconducting quantum circuits, Physics Reports {\bf 718}, 1 (2017).

\bibitem{Painter-19}
M. Mirhosseini, E. Kim, X. Zhang, A. Sipahigil, P. B. Dieterle, A. J. Keller, A. Asenjo-Garcia, D. E. Chang, and O. Painter, Cavity quantum electrodynamics with atom-like mirrors, Nature {\bf 569}, 692 (2019).

\bibitem{Elshaari-20}
A. W. Elshaari, W. Pernice, K. Srinivasan, O. Benson, and V. Zwiller, Hybrid integrated quantum photonic circuits, Nature photonics {\bf 14}, 285 (2020).


\bibitem{Kim-17}
J.-H. Kim, S. Aghaeimeibodi, C. J. K. Richardson, R. P. Leavitt, D. Englund, and E. Waks, Hybrid Integration of Solid-State Quantum Emitters on a Silicon Photonic Chip, Nano Lett. {\bf 17}, 7394 (2017).

\bibitem{Nori-13}
Z.-L. Xiang, S. Ashhab, J. Q. You, and F. Nori, Hybrid quantum circuits: Superconducting circuits interacting with other quantum systems, Rev. Mod. Phys. {\bf 85}, 623 (2013).

\end{thebibliography}
\end{document}